\pdfoutput=1
\documentclass{aa}

\usepackage{graphicx} 
\usepackage[varg]{txfonts}
\usepackage{natbib}
\begin{document} %

   \title{Millimeter radiation from a 3D model of the solar atmosphere}

   \subtitle{I. Diagnosing chromospheric thermal structure}

   \author{M. Loukitcheva
          \inst{1,2}
          \and
          S. K. Solanki \inst{1,3}
           \and
           M. Carlsson  \inst{4}
          \and
          S. M. White \inst{5}
          }

   \institute{Max-Planck-Institut for Sonnensystemforschung, Justus-von-Liebig-Weg 3, 37077 G\"ottingen, Germany\\
    \email{lukicheva@mps.mpg.de}
    \and
    Astronomical Institute, St.Petersburg University, Universitetskii pr. 28, \\ 198504 St.Petersburg, Russia
    \and
    School of Space Research, Kyung Hee University, Yongin, Gyeonggi 446-701, Korea
     \and
    Institute of Theoretical Astrophysics, University of Oslo, P.O. Box 1029, Blindern, N-0315 Oslo, Norway
    \and
    Space Vehicles Directorate, Air Force Research Laboratory, Kirtland AFB, NM, United States
             }

\titlerunning{Millimeter Radiation from a 3D Model}
\authorrunning{M. Loukitcheva et al.}

\date{Received ; accepted}


  \abstract
   {}
   {We use advanced 3D NLTE radiative magnetohydrodynamic simulations of the solar atmosphere to carry out detailed tests of chromospheric diagnostics at millimeter and submillimeter wavelengths.
}
   {We focused on the diagnostics of the thermal structure of the chromosphere in the wavelength bands from 0.4~mm up to 9.6~mm that can be accessed with the Atacama
Large Millimeter/Submillimeter Array (ALMA) and investigated how these diagnostics are affected by the instrumental resolution. }
   {We find that the formation height range of the millimeter radiation depends on the location in the simulation domain and is related to the underlying magnetic structure. Nonetheless, the brightness temperature is a reasonable measure of the gas temperature
at the effective formation height at a given location on the solar surface. There is considerable scatter in this relationship, but this is significantly reduced when very weak magnetic fields
are avoided. Our results indicate that although instrumental smearing
reduces the correlation between brightness and temperature, millimeter brightness can still be
used to reliably diagnose electron temperature up to a resolution of 1\arcsec. If the resolution is more degraded, then the value of the diagnostic diminishes rapidly.
}
   {We conclude that millimeter brightness can image the chromospheric thermal structure
at the height at which the radiation is formed. Thus multiwavelength observations with
ALMA with a narrow step in wavelength should provide sufficient information for a tomographic imaging of the
chromosphere.}

   \keywords{Sun: atmosphere -- Sun: chromosphere -- Sun: radio radiation -- Sun: magnetic fields
               }

   \maketitle

\section{Introduction}

 As a result of its particularly complicated physics, the chromosphere remains the least understood layer of the solar atmosphere. The past decade of modeling efforts has enormously advanced our understanding of this enigmatic layer, however. In recent years, 3D modeling of the chromosphere, including much of the necessary physics, has finally started to become feasible. Advanced 3D magneto-hydrodynamic models run from the upper convection zone through the chromosphere into the lower corona \citep{gudiksen} and take into account physical effects such as magnetic field dynamics, thermal conduction, ambipolar diffusion, and the Hall effect, non-equilibrium ionization, and NLTE effects \citep[see, e.g.,][for a review]{carlsson09}. These simulations provide the opportunity to model the formation of various chromospheric lines and continua and to study the diagnostic potential of chromospheric data \citep{leenaarts13a,leenaarts13b,pereira,stepan,leenaarts12}.

 Chromospheric diagnostics, such as Ca {\sc ii} H\&K, H$\alpha$, Ca {\sc ii} 854.2 nm,  and Mg {\sc ii} h\&k lines, suffer from the fact that they are formed out of local thermodynamic equilibrium (LTE) and thus decouple from the local conditions. In addition, their formation is complex. An alternative chromospheric diagnostic is provided by observations of the radio continuum at submillimeter and millimeter wavelengths \citep{loukitcheva}. The intensities of submm/mm continua, whose source function can be treated in LTE \citep{Rutten}, depend linearly on temperature and therefore may be able to provide a sensitive test of numerical models. The additional advantage of submm/mm continua is that their intensities can be easily synthesized from the models.

 \citet{loukitcheva} compared a large collection of submillimeter and millimeter observations with synthetic brightness temperatures calculated in classical standard models by \citet{fontenla} and the 1D dynamic model of \citet{carlsson95} and demonstrated that both the classical and the dynamic model provide a reasonable fit to observed temporally and spatially averaged millimeter data. The analysis of the 1D dynamic simulations of Carlsson and Stein further revealed that radio emission at millimeter wavelengths is extremely sensitive to dynamic processes in the chromosphere if these are spatially and temporally resolved \citep{loukitcheva,loukitcheva2006}.

\begin{figure*}[!htb]
   \centering
   \includegraphics[width=0.8\textwidth]{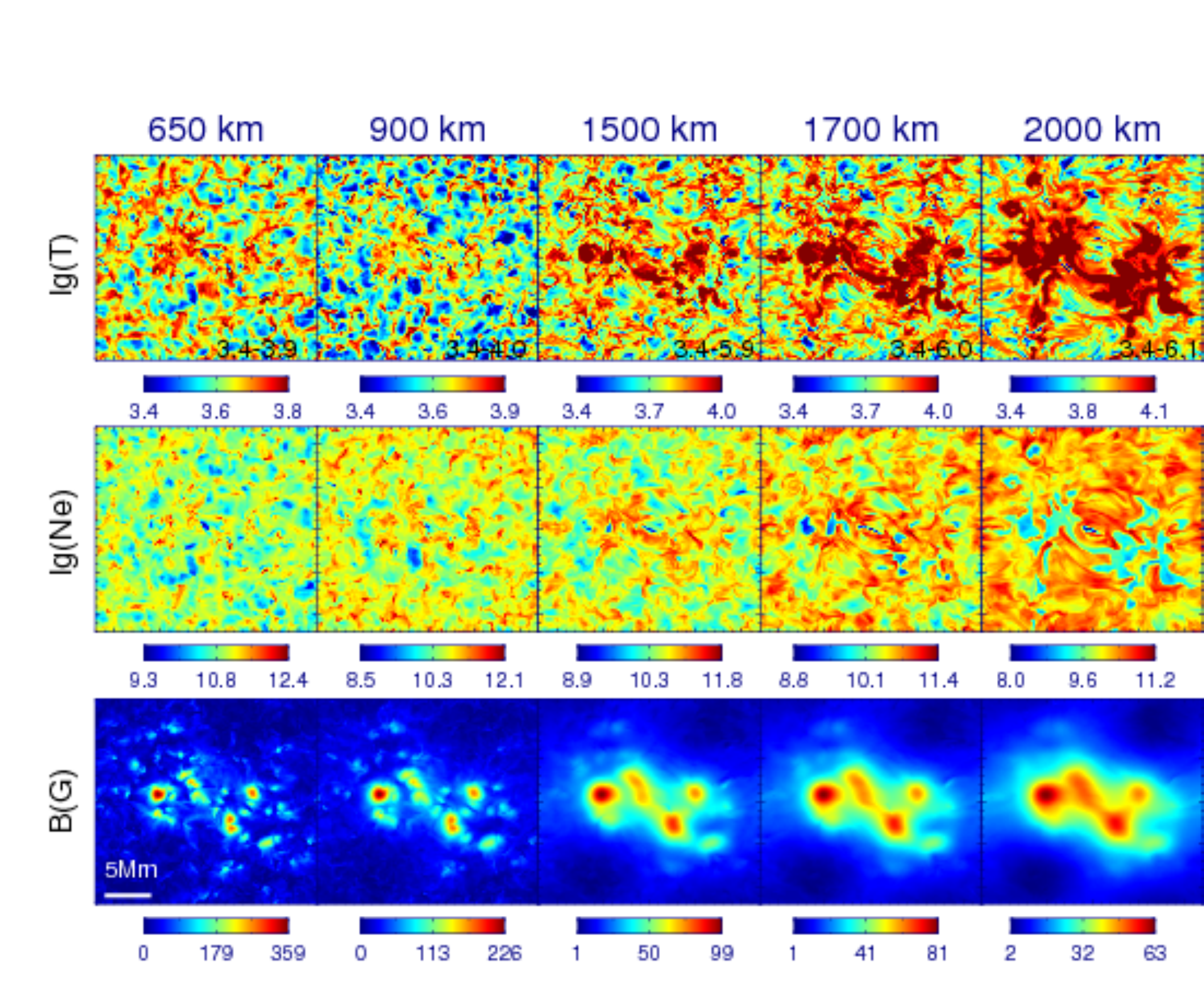}
      \caption{Horizontal cuts through a 3D MHD simulation snapshot. Plotted are the electron temperature (top panel), electron number density (middle panel), and magnetic field strength (bottom) at geometrical heights of 650, 900, 1500, 1700, and 2000 km. To enhance the temperature contrast, the displayed temperature range is set to that of the synthetic brightness maps shown in Fig.~\ref{fig2}. The range of temperatures is given in the lower right corner of the frames. The field size is 24~Mm x 24~Mm.}
         \label{fig1}
   \end{figure*}

\citet{wed07} were the first to study submm/mm brightness synthesized from 3D radiation
(magneto-)hydrodynamic simulations, using those of \citet{wed04} and \citet{leenaarts06}. These 3D (M)HD models were constructed under the assumptions of LTE for the radiative transfer and the equation of state to make the computations feasible. In addition, \citet{wed07} studied submm/mm brightness from representative model snapshots with non-equilibrium electron densities and a
weak magnetic field.  The authors reported mm brightness maps with filamentary brightenings, resulting from shock-induced thermal structure, and fainter regions in between. The average brightness temperature and relative contrast were found to increase with wavelength.

This paper is the first in a series in which we use advanced 3D radiative magnetohydrodynamic simulations of the solar atmosphere performed with the Bifrost code \citep{gudiksen} to carry out detailed tests of chromospheric diagnostics at millimeter and submillimeter wavelengths.
In this paper we focus on the diagnostics of the chromospheric thermal structure in the wavelength bands accessible to the advanced radio interferometer Atacama Large Millimeter/Submillimeter Array (ALMA). Located on the Chilean Chajnantor plateau above 5000~m altitude, ALMA consists of a giant array of 12 m antennas (the 12 m array) with baselines of up to 16 km, which aims at achieving high spatial resolution, and an additional compact array of 7
m and 12 m antennas that greatly enhance ALMA's ability to image extended targets. At the time of writing, ALMA observes at wavelengths in the range 3.6~mm to 0.4~mm (84 to 720 GHz) in dual polarization mode with prospects of going to longer (up to 10~mm) and shorter (down to 0.3~mm) wavelengths in the future. We simulate the mm/submm brightness at ALMA wavelengths from the 3D MHD simulations and compare the results with the actual temperature distributions to assess the accuracy of the mm-wavelength diagnostics.

The structure of the paper is as follows: In Sects.~2 and 3 we describe the 3D model atmosphere based on the Bifrost code and the submm/mm radiative transfer computations. The analysis of the synthetic brightness is presented and discussed in Sect.~4. In this section we also investigate the effect of spatial smearing of the model brightness in the light of future interferometric observations with ALMA. We summarize our findings and conclude in Sect.~5.

\section{3D Model atmosphere}

We studied the formation of millimeter and submillimeter continua in a snapshot of a 3D radiation-MHD simulation performed with the Bifrost code \citep{gudiksen}. We used snapshot 385 of the simulation ``en024048{\_}hion'' made available through the Interface Region Imaging Spectrograph \citep[IRIS,][]{De Pontie} project at the Hinode Science Data Centre Europe (http://sdc.uio.no). The same snapshot was used by \citet{leenaarts12} to
investigate the H$\alpha$ line formation and by \citet{rodriguez} to study Ca~{\sc ii} 8542 spectra. It was also used by \citet{leenaarts13a} and other papers in that series to investigate the formation of the Mg~{\sc ii} h and k lines in the solar atmosphere. Bifrost solves the equations of resistive MHD on a staggered Cartesian grid. The simulation includes optically thick radiative transfer in the photosphere and low chromosphere, parameterized radiative losses in the upper chromosphere, transition region, and corona \citep{carlsson12}, thermal conduction along magnetic field lines \citep{gudiksen}, and an equation of state that includes the effects of non-equilibrium ionization of hydrogen \citep{leenaarts07}.

The simulation covers a physical extent of 24~x~24~x~16.8 Mm, with a grid of 504~x~504~x~496 cells, extending from 2.4~Mm below the average height of $\tau_{500} = 1$, which corresponds to $Z=0$, to 14.4~Mm above, so that it covers the upper convection zone, photosphere, chromosphere, and the lower corona. The horizontal axes have an equidistant grid spacing of 48~km (0.06\arcsec), the vertical grid spacing is nonuniform, with a spacing of 19~km between $Z = -1$~Mm and $Z = 5$~Mm. The spacing increases toward the bottom and top of the computational domain to a maximum of 98 km. The simulation contains a magnetic field with an average unsigned strength of 50~G in the photosphere, concentrated in the photosphere in two clusters of opposite polarity lying 8~Mm apart, representing two patches of quiet-Sun network. We refer to the snapshot from this simulation we analyzed simply as the 3D MHD snapshot. For more details of this snapshot we refer to \citet{carlsson13,carlsson14}.

In Fig.~\ref{fig1} we show examples of horizontal slices through the 3D MHD snapshot of electron temperature (upper panel), electron number density (middle panel), and magnetic field strength (bottom) taken at the geometrical heights $Z=650$, $900$, $1500$, $1700,$ and $2000$~km. These heights very roughly correspond to the effective formation heights of the radiation at 0.4, 1.0, 3.0, 4.5, and 10~mm, respectively (see Sect.~\ref{sectCF}).
Between the magnetic patches the temperature shows filamentary structure that follows the general orientation of the magnetic field. A hot chromospheric canopy lies above and between the photospheric magnetic field concentrations, and coronal temperatures are reached already at heights of around 1700~km. The electron density displays a similarly complex pattern, with a clear anticorrelation with the temperature above magnetic patches at heights above 1700~km. The magnetic field concentrations expand with height from their photospheric footpoints (with a maximum field strength of 2200~G) to a volume-filling field higher up with a maximum strength of 60~G at a geometrical height of 2000~km (see also Fig.~\ref{fig3}).

 \begin{figure*}[!htb]
  \centering
     \includegraphics[width=0.78\textwidth]{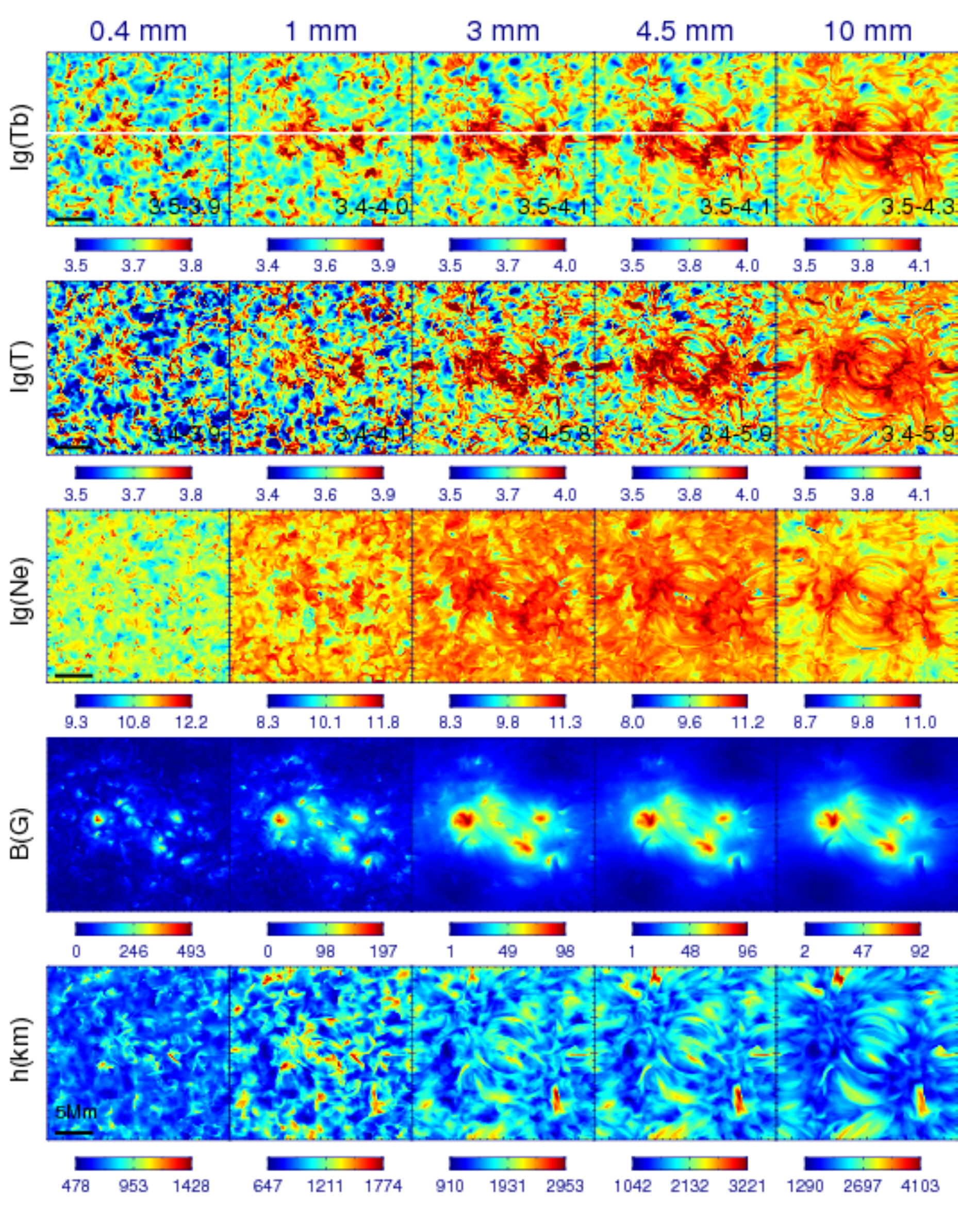}
      \caption{Maps of simulated mm brightness at the resolution of the model for the wavelengths 0.4, 1.0, 3.0, 4.5, and 10~mm (labeled in the top row). The      horizontal white line in the top row marks the position of the vertical cut at Y=0.9~Mm displayed in Fig.~\ref{fig3}. We also plot the effective formation heights of the mm radiation (bottom row) and the corresponding electron temperatures, electron number densities, and magnetic field strength taken at the effective formation heights (three middle rows). The values of the most extreme 1\% of points are clipped in the brightness maps. The displayed temperature range for electron temperatures is set to that of the mm brightness maps. The range of covered temperatures is given near the bottom of the frames. The field size is 24~Mm x 24~Mm.
              }
         \label{fig2}
   \end{figure*}

\section{Synthesis of mm brightness}

We calculated the submillimeter and millimeter radiation  emerging from the 3D MHD snapshot at 32 wavelengths from 0.4~mm to 10~mm. In the calculations we assumed each vertical column in the 3D snapshot to be an independent 1D plane-parallel atmosphere. The radiative transfer was computed under the assumption that bremsstrahlung opacity is responsible for the mm continuum radiation in the quiet Sun. We distinguished two sources
of opacity: opacity due to encounters between free electrons and protons, and opacity due to encounters between free electrons and neutral hydrogen. They are
commonly referred to as H$^0$ opacity and H$^-$ opacity. The corresponding absorption coefficients depend on the effective
number of collisions undergone by an electron per unit time, including collisions with protons ($\chi_\mathrm{ep}\propto \frac{N_\mathrm{e} N_\mathrm{p}}
{T_\mathrm{e}^{3/2} \nu^{2}}$) and with neutral hydrogen atoms ($\chi_\mathrm{eH}\propto \frac{N_\mathrm{e}
\,N_\mathrm{H}\,T_\mathrm{e}^{1/2}}{\nu^{2}}$). Expressions for calculating the radiative transfer were taken from \citet{zheleznyakov} and can also be found in \citet{loukitcheva}. In the presence of a magnetic field in quasilongitudinal approximation, the absorption coefficients for extraordinary ($e$) and ordinary ($o$) modes are determined by the relation \citep{zlotnik}
  \begin{equation}
   \chi_\mathrm{e,o} \simeq \frac{\chi_\mathrm{e,o}^{0}}{1\mp\sqrt{\frac{\omega_B}{\omega}}|\cos\alpha|}
     \label{eq1}
   ,\end{equation}
   where $\chi_\mathrm{e,o}^{0}$ characterizes absorption ($\chi_\mathrm{ep}$  or $\chi_\mathrm{eH}$) in the absence of the magnetic field, $\omega$ is the observing frequency, $\omega_B$ is the gyrofrequency, $\alpha$ is the angle between the magnetic field and the line of sight, the minus sign corresponds to the $e$-mode, and the plus sign to the $o$-mode. For a weak magnetic field ($\mid \sqrt{\frac{\omega_B}{\omega}} \cos \alpha \mid \ll 1$) the quasilongitudinal approximation is valid for all angles between the magnetic field and the line of sight, except for narrow intervals of angles close to transverse propagation \citep{zheleznyakov}.

We considered the radiative transfer equation in terms of brightness temperature $T_\mathrm{b}^{e,o}$ separately for the $e$- and $o$-mode,
   \begin{equation}
   \frac{dT_\mathrm{b}^{e,o}}{d\tau} =
    T_\mathrm{e} - T_\mathrm{b}^{e,o}
              ,   \label{eq2}
   \end{equation}
   where $T_\mathrm{e}=T_\mathrm{e}(l)$ is the profile of the kinetic temperature
   along the light path, $\tau(l) = \int^l_{l_\mathrm{0}} \chi_\mathrm{e,o}(l)dl$
   is the optical depth,  $l$ is geometrical
   distance along the light path,  and $\chi_\mathrm{e,o}$ is the
   corresponding absorption coefficient.
   The solution of the radiative transfer equation on a discrete grid can be
   expressed in terms of $T_\mathrm{e}$ as \citep{hagen}
   \begin{equation}
   T_\mathrm{b}^{e,o}=\sum_{r=1}^n(1-\exp^{-\chi_\mathrm{e,o}^{r}\Delta
   h})\,T_\mathrm{e}^r \exp^{-\sum_{s=1}^{r-1} \chi_\mathrm{s}\Delta h}
   ,      \label{eq6}
   \end{equation}
   where $\chi_\mathrm{e,o}^{r}$ stands for $\chi_\mathrm{e}$ or $\chi_\mathrm{o}$ at position $r$, $r=1$ corresponds to the layer  at the top of the
   atmosphere, $\Delta h$ is the grid distance.
   The items within the sum represent the contribution of various layers to the emerging radiation intensity. We
   refer to them as the values of the (unnormalized) contribution function (CF) to $T_\mathrm{b}^{e,o}$.
   The total brightness $T_\mathrm{b}$ and the degree of circular polarization $P$
   at wavelength $\lambda$ are defined as
   \begin{equation}
    T_\mathrm{b}^{\lambda}=\frac{T_\mathrm{b}^{e}+T_\mathrm{b}^{o}}{2},
\\
      P=\frac{T_\mathrm{b}^{e}-T_\mathrm{b}^{o}}{T_\mathrm{b}^{e}+T_\mathrm{b}^{o}}
   .\end{equation}
   Hereafter we discuss the total brightness at millimeter wavelengths as a primary diagnostics of the chromospheric thermal structure, leaving the discussion of the circular polarization for the next paper of the series, which will be on the subject of the chromospheric magnetic field.

\section{Results}
\subsection{Distribution of brightness temperature}

In Fig.~\ref{fig2} we show the results of the brightness calculations for 0.4~mm (ALMA band~9), 1~mm (band~7), 3~mm (band~3), 4.5~mm (band~2) and 10~mm (band~1). The brightness temperature at all considered wavelengths exhibits a complex pattern of intermittent bright and dark regions (top row of Fig.~\ref{fig2}) similar to those seen in the electron temperature and density horizontal cuts through the model atmosphere (see Fig.~\ref{fig1}). In the central part of the field of view the most prominent features are bright elongated fibrils that extend outward from the network patches and are aligned along the chromospheric magnetic field lines. At the shortest wavelengths the imprint of the central loop-like magnetically formed structures is weak, but it becomes more pronounced toward longer wavelengths.

\begin{table}
\caption{Lowest and highest brightness temperature, average brightness temperature $<T_b>$, RMS variation $T_b^{rms}$, and relative brightness temperature contrast $\frac{T_b^{rms}}{<T_b>}$ for a number of analyzed wavelengths.}             
\label{table:1}      
\centering                          
\begin{tabular}{l c c c c c}        
\hline\hline                 
\noalign{\smallskip}
$\lambda$, mm & $T_{b}^{min}$, K & $T_{b}^{max}$, K & $<T_b>$, K & $T_b^{rms}$, K & $\frac{T_b^{rms}}{<T_b>}$ \\    
\noalign{\smallskip}
\hline                        
\noalign{\smallskip}
0.4 & 3336   &   8446 & 4507   &   503  &   0.11\\
1.0 & 2706   &   11094 & 4786   &   823  &   0.17\\
3.0 & 2945   &   13104 & 6090   &   1333 &   0.22\\
3.6 & 3101   &   13227 & 6394   &   1402  &   0.22\\
4.5 & 3054    &  13637 & 6786   &   1485  &   0.22\\
10.0 & 2924   &  20082 & 8615   &   1788  &   0.21\\
\noalign{\smallskip}
\hline                                   
\end{tabular}
\end{table}

\begin{figure}[!htb]
  \centering
            \includegraphics[trim=100 0 0 100,clip, width=0.55\textwidth]{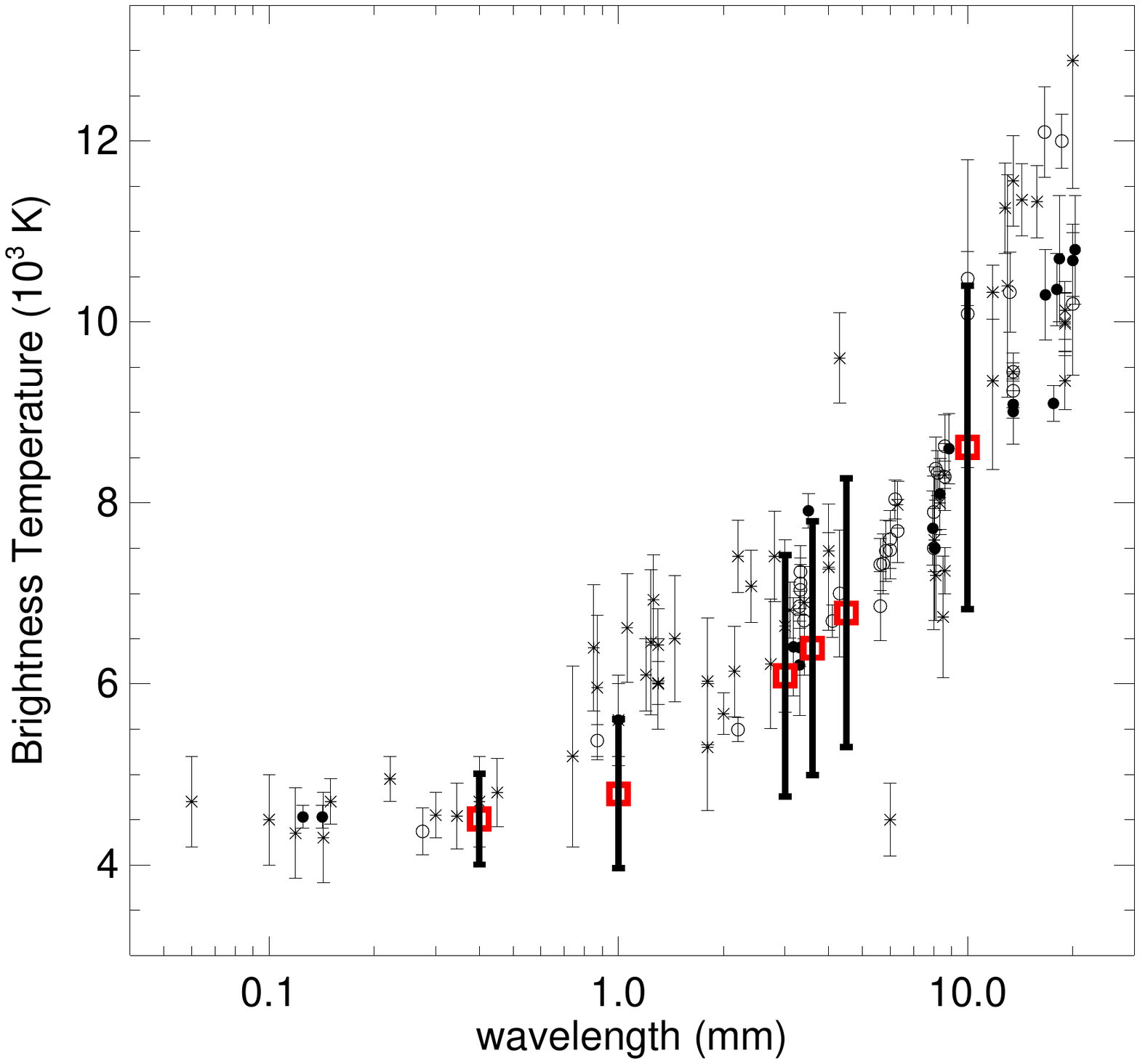}
      \caption{Simulated millimeter brightness (red squares) overlaid on the observed brightness temperatures from \citet{loukitcheva}. Data obtained around solar cycle minimum, solar maximum, and at intermediate phases are depicted by filled circles, open circles, and stars, respectively.
              }
         \label{fig3a}
   \end{figure}

To quantify the brightness distributions at mm wavelengths, we list in Table~\ref{table:1} some statistics including the lowest and highest brightness temperature, average brightness temperature $<T_b>$, RMS variation ${T}_b^{rms}$, and the relative brightness temperature contrast $\frac{{T}_b^{rms}}{<T_b>}$ for a number of analyzed wavelengths. The average brightness temperature as well as the highest brightness increase with wavelength as a
result of the shifting contributing heights to higher (hotter) layers in the atmosphere (see Figs.~\ref{fig5}, \ref{fig6a}, \ref{fig6}, \ref{fig7}, \ref{fig8}, and \ref{fig9}, which are discussed in detail later). The RMS variation $T_b^{rms}$ also steadily increases with $\lambda$. A more complex behavior is displayed by the relative brightness contrast, which initially increases with wavelength, reaches a value of $0.22$ at around $\lambda$=3~mm, and stays almost constant at longer wavelengths. In Fig.~\ref{fig3a}  we plot average brightness temperatures $<T_b>$, with ${T}_b^{rms}$ as error bars, overlaid on the observed millimeter brightness spectra from the compilation made by \citet{loukitcheva}. Gratifyingly, $<T_b>$ calculated from the snapshot of the 3D MHD model (red squares in Fig.~\ref{fig3a}) are consistent within the error bars with the available observations of mm brightness. The only small discrepancy occurs at 1~mm, where, in spite of some overlap, the simulated $T_b$ appears to lie somewhat below the observational data. Longward of 3~mm the simulated values are significantly higher than those reported in \citet{wed07}.

 \begin{figure}[!htb]
  \centering
            \includegraphics[trim=0 0 100 100,clip,width=0.45\textwidth]{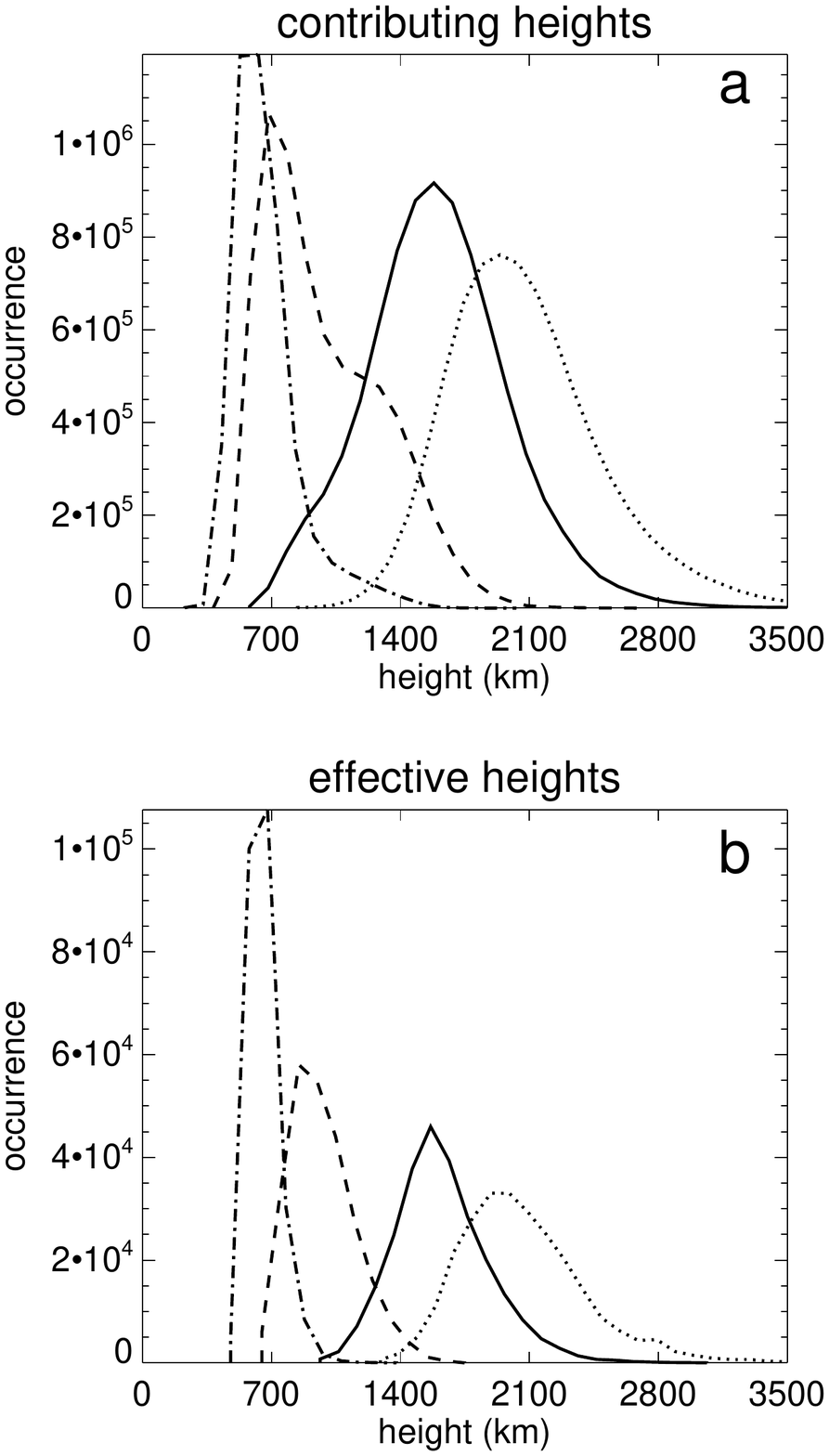}
      \caption{\textbf{(a)} Histograms of the heights contributing more than 1\% of the total contribution to the emerging intensity at 0.4~mm (dot-dashed), 1~mm (dashed), 3.6~mm (solid), and 10~mm (dotted). \textbf{(b)} The same for the effective formation heights.
              }
         \label{fig51}
 \end{figure}

\subsection{Effective formation heights}\label{sectCF}

The bottom row of Fig.~\ref{fig2} shows maps of the effective formation height for five wavelengths. We define the effective formation height as the height corresponding to the centroid of the contribution function.
At 0.4~mm the radiation originates mainly in layers close to the traditional temperature minimum region and lower chromosphere with a few, very localized excursions to heights above 1000~km. The average formation height of this wavelength is around 650~km. With increasing wavelengths, we clearly sample higher chromospheric layers. Effective formation heights, averaged over all spatial locations, reach values of about 900~km for 1~mm, 1500~km for 3~mm, 1700~km for 4.5~mm, and around 2000~km for 10~mm. In the expanding magnetic patches that overlie photospheric magnetic footpoints and show enhanced temperature, we normally see down to lower chromospheric heights (e.g., 1200-1500~km at 3~mm), while in loop-like structures that connect the magnetic concentrations, we sample higher layers (1500-2000~km at 3~mm). Details on the formation heights of individual brightness features can be found in Sects.~\ref{sectXZ} and ~\ref{sectXYZ}.

To check how representative of the true formation layers the effective height is, we plot in the second top row of Fig.~\ref{fig2} the electron temperature $T_e$ taken at the effective formation
heights. For completeness, we show in the third and fourth rows of Fig.~\ref{fig2} the same for the electron number density $n_e$ and the magnetic field strength $B$. Qualitatively, the properties of the atmosphere taken at the effective formation heights (Fig.~\ref{fig2}) look quite similar to the atmospheric properties of the layers corresponding to the average effective heights (Fig.~\ref{fig1}), but there is a significant difference in the range of values, specifically for the magnetic field. From the comparison of the top row of Fig.~\ref{fig1}, which shows $T_e$ at the average formation heights of mm radiation in the 3D MHD snapshot, with
the two upper rows of Fig.~\ref{fig2}, which depict synthetic $T_b$ and $T_e$ at the effective formation heights, we conclude that in general, mm brightness can serve as a measure of temperature in the solar atmosphere, albeit not a
perfect measure. We report the results of a correlation analysis
that provides a more quantitative comparison in Sect.~\ref{sectCor}.

 \begin{figure}[!htb]
  \centering
            \includegraphics[trim=300 0 0 0,clip, width=0.45\textwidth]{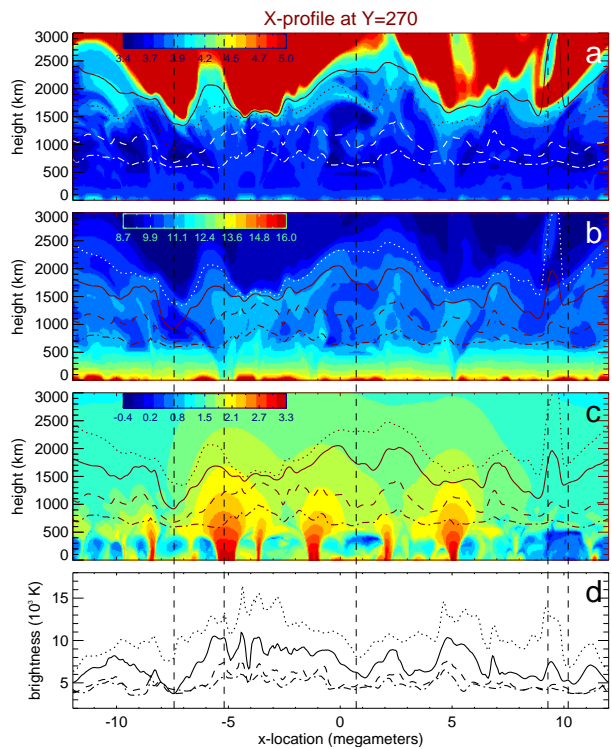}
      \caption{XZ-profile of the 3D MHD snapshot along the cut at Y=0.9~Mm, marked by a horizontal line in Fig.~\ref{fig2}. (a) Electron temperature on a logarithmic scale clipped at $10^5$~K to enhance the contrast at lower heights. (b) Electron number density on a logarithmic scale. (c) Magnetic field strength on a logarithmic scale. Overplotted dot-dashed, dashed, solid, and dotted lines in panels (a)-(c) depict the effective formation heights for the radiation at 0.4~mm, 1.0~mm, 3.6~mm, and 10~mm, respectively. (d) Brightness temperature at the wavelengths 0.4~mm (dot-dashed), 1.0~mm (dashed), 3.6~mm (solid), and 10~mm (dotted). In panels (a)-(c) the y-axis scale is expanded. In all panels vertical dashed lines mark the locations of the different structures presented in Figs.~\ref{fig6a}, \ref{fig6}, \ref{fig7}, \ref{fig8}, and \ref{fig9}.
              }
         \label{fig3}
   \end{figure}

Statistical distributions of the heights that contribute to the emerging intensity are presented in Fig.~\ref{fig51} in the form of histograms for four of the analyzed wavelengths together with the distributions of the effective formation heights for comparison.
The corresponding histogram peaks in the two panels coincide with an accuracy of 50~km for all wavelengths except for that
at 1~mm. At 1~mm, the discrepancy between the peaks in the contributing and effective height histograms reaches 160~km, and the histograms differ in form, which suggests that statistically we need to be more careful at 1 mm in interpreting the effective formation
heights.  

\subsection{Analysis of a vertical slice through the snapshot}\label{sectXZ}

In Fig.~\ref{fig3} we provide an example of an XZ-slice through the atmosphere along the cut at Y=0.9~Mm marked in the top row of Fig.~\ref{fig2} by the horizontal white line. The panels (a-c) of Fig.~\ref{fig3} display the electron temperature $T_e$, the electron number density $n_e$ , and magnetic field strength $B$ along the cut, with the effective formation heights at four selected wavelengths overplotted. The bottom panel (d) shows the synthesized brightness along the cut at $\lambda$=0.4~mm (dot-dashed), 1.0~mm (dashed), 3.6~mm (solid), and 10~mm (dotted).

To some extent, the effective formation heights of mm radiation replicate the distribution of the local spatial inhomogeneities in electron temperature and number density along the cut. To a larger extent, however, they are defined by the temperature and number density structure along the line of sight (in particular, the strong local gradients of temperature and density that influence opacity peaks and therefore the form of the contribution function) and thus mix contributions from both high and low layers.

    \begin{figure}[!htb]
  \centering
            \includegraphics[trim=0 0 150 0,clip,width=0.45\textwidth]{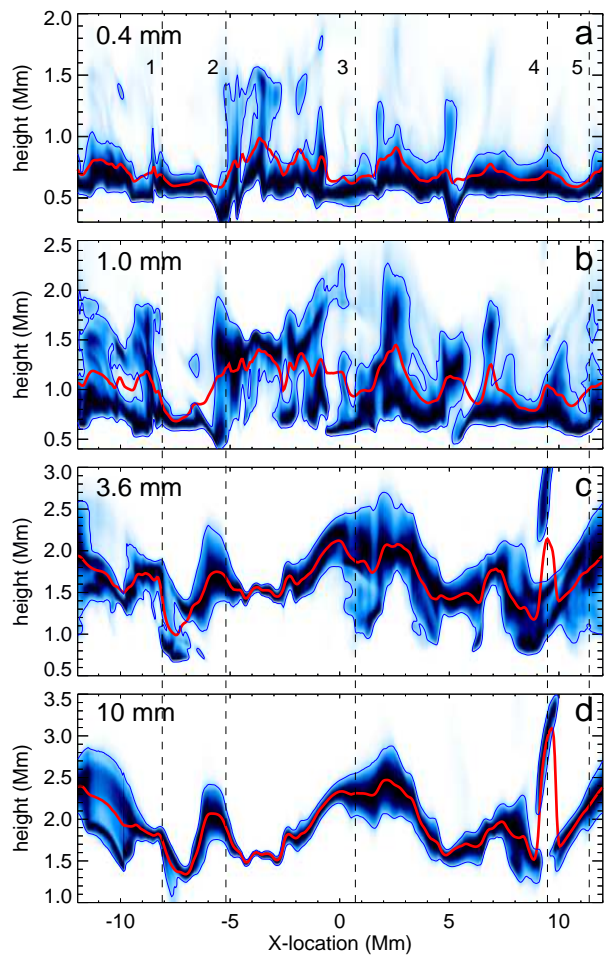}
      \caption{Gray-scale figures of the contribution function along the cut at Y $=0.9$ displayed in Fig.~\ref{fig3}. Dark color corresponds to higher amplitudes. Contribution functions in each column are normalized to the highest value of the column. Heights with contribution function amplitudes equal to 10\% of the highest contribution to the emerging intensity are marked with blue contours. Red solid lines represent effective formation heights. The various panels show results for 0.4~mm (a), 1~mm (b), 3.6~mm (c), and 10~mm (d). In all panels the y-axis scale is expanded. The dashed lines are the same as in Fig.~\ref{fig3}.
              }
         \label{fig5}
   \end{figure}

Away from the regions with a strong magnetic field, the effective formation height tends to be
lower than in the vicinity of magnetic field concentrations (see Fig.~\ref{fig3}c), although there are exceptions. Directly above the field concentrations we see a relative shift of the iso-Kelvin temperature curves (and consequently of the effective formation heights) to higher layers than the surrounding locations. Synthesized brightness at mm wavelengths does not necessarily follow the effective formation height profile, but there is a tendency for local brightenings above the field concentrations (see curves of the same type in Fig.~\ref{fig3}a-c and \ref{fig3}d).

  \begin{figure*}[!htb]
  \centering
           \includegraphics[trim=0 0 0 350,clip,width=0.95\textwidth]{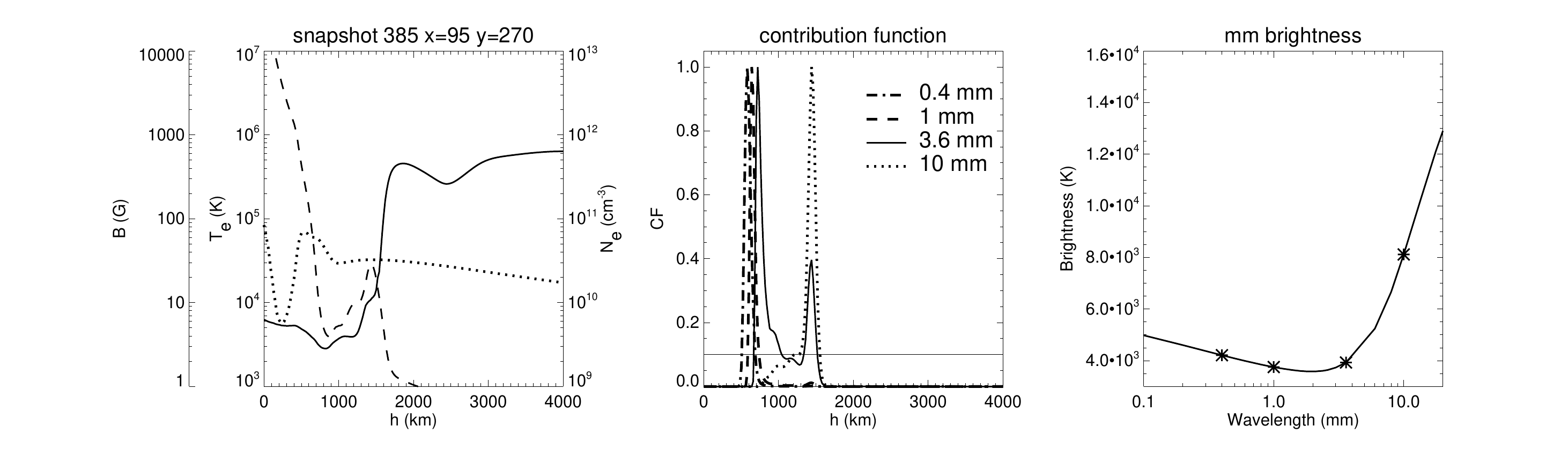}
      \caption{Spatial location (x,y)=(-8.1,0.9)~Mm, marked ''1'' in Fig.~\ref{fig5}. (a) Electron temperature (solid curve), electron number density (dashed curve), and magnetic field strength (dotted) as a function of height. (b) Contribution functions to emerging intensity (normalized to their maximum) as labeled in the figure. The horizontal line marks 10\% of the highest value for comparison with Fig.~\ref{fig5}. (c) Resulting brightness spectrum. Stars mark the four wavelengths analyzed in detail.
              }
         \label{fig6a}
   \end{figure*}

 \begin{figure*}[!htb]
  \centering
           \includegraphics[trim=0 0 0 350,clip,width=0.95\textwidth]{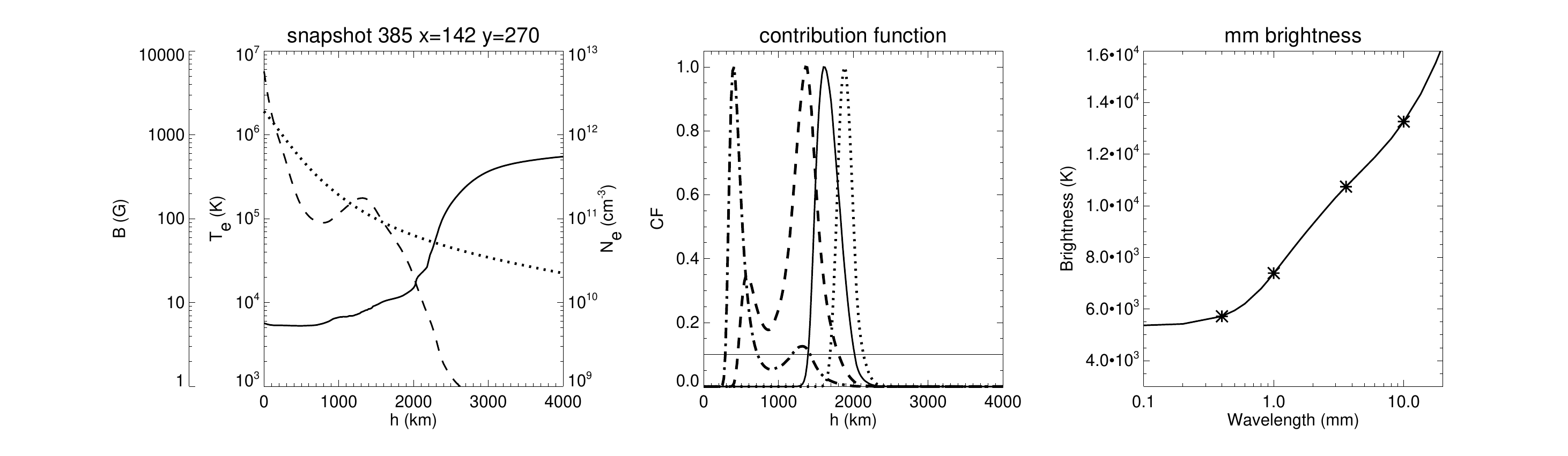}
      \caption{Same as in Fig.~\ref{fig6a} for the spatial location (-5.2,0.9)~Mm, marked ''2'' in Fig.~\ref{fig5}.
              }
         \label{fig6}
   \end{figure*}

 \begin{figure*}[!htb]
  \centering
            \includegraphics[trim=0 0 0 350,clip,width=0.95\textwidth]{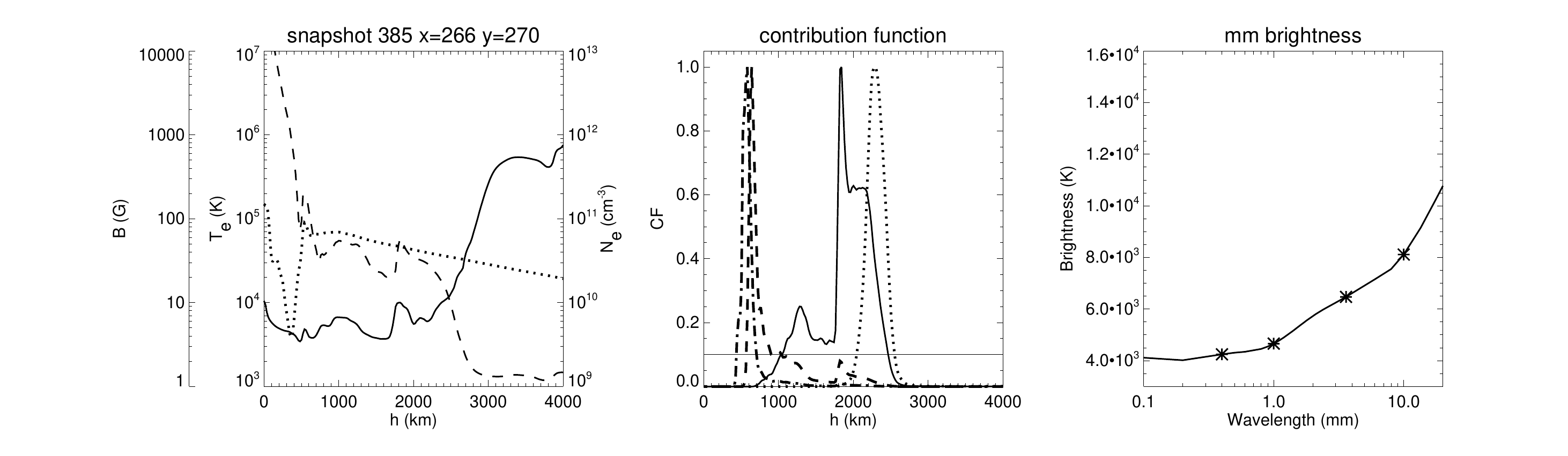}
      \caption{Same as in Fig.~\ref{fig6a} for the spatial location (0.7,0.9)~Mm, marked ''3'' in Fig.~\ref{fig5}.
              }
         \label{fig7}
   \end{figure*}

 \begin{figure*}[!htb]
  \centering
            \includegraphics[trim=0 0 0 350,clip,width=0.95\textwidth]{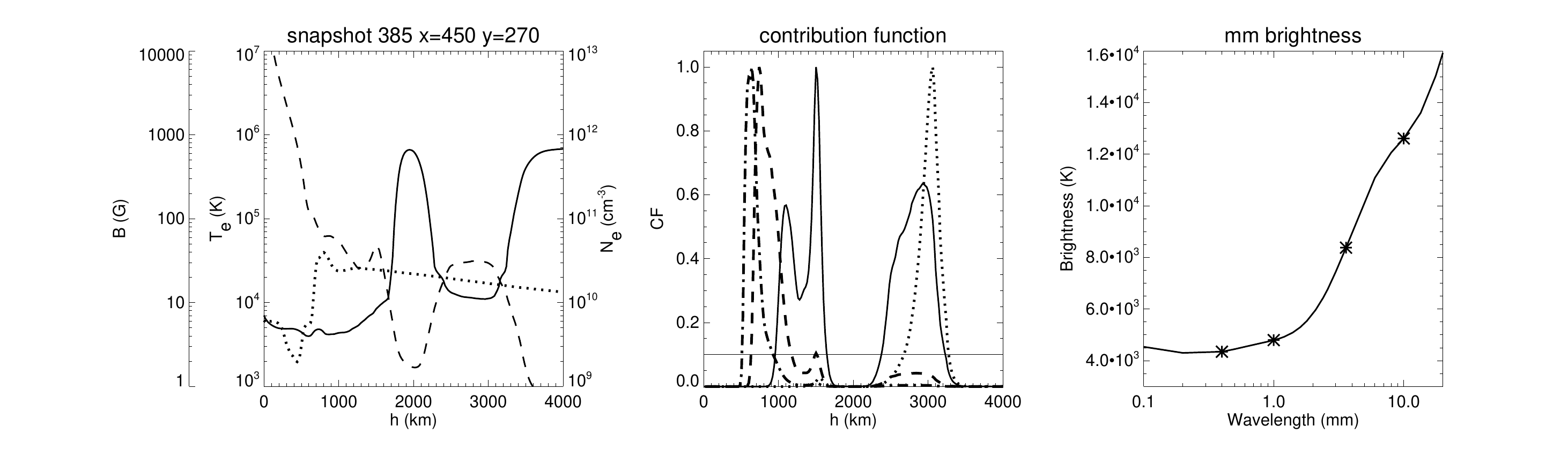}
      \caption{Same as in Fig.~\ref{fig6a} for the spatial location (9.5,0.9)~Mm, marked ''4'' in Fig.~\ref{fig5}.
              }
         \label{fig8}
   \end{figure*}
 \begin{figure*}[!htb]
  \centering
              \includegraphics[trim=0 0 0 350,clip,width=0.95\textwidth]{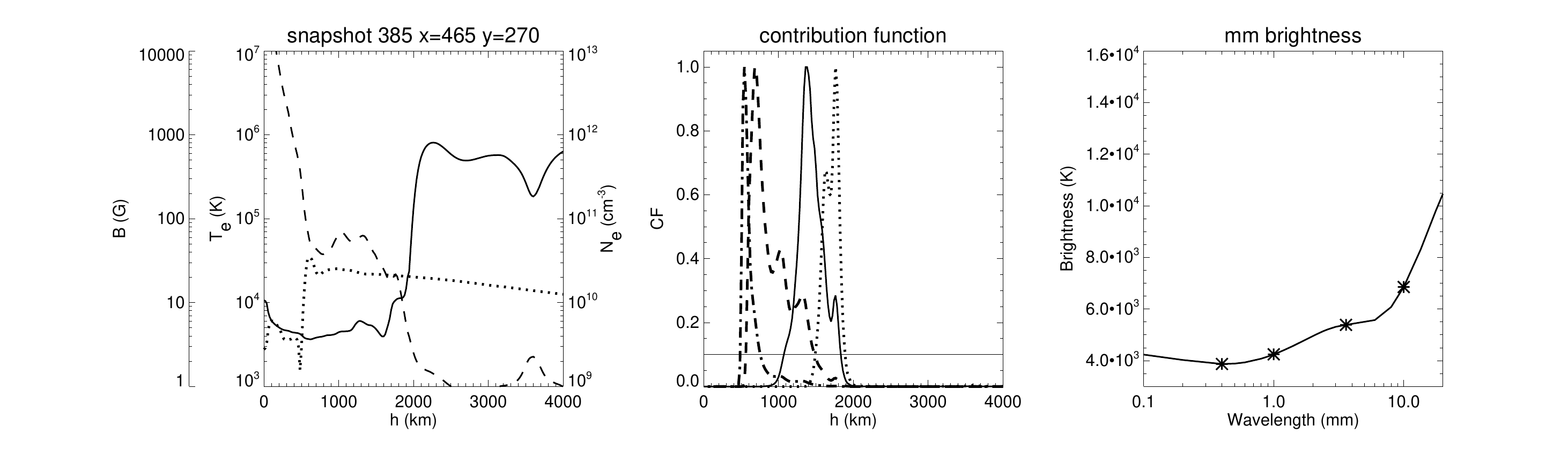}
      \caption{Same as in Fig.~\ref{fig6a} for the spatial location (10.2,0.9)~Mm, marked ''5'' in Fig.~\ref{fig5}.
              }
         \label{fig9}
   \end{figure*}

In Fig.~\ref{fig5} we show the range of geometrical heights that contribute to the emerging intensity as a function of location along the  vertical slice from Fig.~\ref{fig3}. The heights marked in Fig.~\ref{fig5} with blue contours at a level of 10\% of the CF maximum can serve as a measure of the height range sampled by each of the four analyzed wavelengths. For comparison, we plot the corresponding effective formation height (red solid line) at each wavelength.
Owing to the complex thermal structure, a broad height range is involved in emitting radiation at all considered wavelengths, resulting in quite complicated contribution functions (see Sect.~\ref{sectXYZ} for details). The figure demonstrates the limitations of using effective heights to assign the synthesized brightness to the thermal structure of the atmosphere. At a number of locations the contribution achieves significant values over two distinct height ranges with double-peaked contribution functions (see, e.g., $\lambda$=1~mm in Fig.~\ref{fig6} and $\lambda$=3.6~mm in Figs.~\ref{fig7} and \ref{fig8}).

\subsection{Analysis of individual profiles}\label{sectXYZ}

In this section, we analyze the formation of the mm brightness spectra at five locations along the XZ-slice that represent different features in the solar atmosphere and display a broad range of behavior. In Fig.~\ref{fig3} these locations are marked by dashed vertical lines, and in the top panel of Fig.~\ref{fig5} they
are additionally labeled with numbers from $1$ to $5$. The properties of the spatially resolved atmospheres (electron temperature, electron number density, magnetic field strength) as a function of geometrical height, contribution functions at four wavelengths and corresponding brightness spectra are shown in Figs.~\ref{fig6a}, \ref{fig6}, \ref{fig7}, \ref{fig8}, and \ref{fig9}, for features 1 to 5. Note that for better comparison the scales remain unchanged from one figure to the next.

\begin{itemize}
  \item The spatial location $(x,y)=(-8.1,0.9)$~Mm (marked  "1" in Fig.~\ref{fig5}) is detailed in Fig.~\ref{fig6a}: At the solar surface the location corresponds to an intergranular feature with a rather weak magnetic field ($<100$~G). Its temperature distribution is characterized by a low-lying transition region (at 1500~km). The electron density manifests a localized increase (by an order of magnitude) at around 1400~km. The contribution functions at $0.4$~mm and $1$~mm are very sharply defined, and the contributing formation heights are in the range $600-700$~km, where temperature decreases with height, leading to a temperature minimum at around $800$~km. At $\lambda$=3.6~mm the contribution function is double peaked, with the main peak at lower height covering the extended temperature minimum region, and the narrow, weak secondary peak in the chromosphere and lower transition region. As a consequence, the synthesized brightnesses are extremely low (around 4000~K) and brightness temperatures at 1~mm and 3.6~mm are lower than that at 0.4~mm. Radiation at 10~mm is formed over a range of heights with a sharp and clear maximum contribution from around $1450$~km. The resulting brightnesses are also significantly lower than the average values for this wavelength.

  \item Spatial location $(x,y)=(-5.2,0.9)$~Mm (Fig.~\ref{fig6}, marked ''2'' in Fig.~\ref{fig5}): Here $B$ reaches 1900~G in the photospheric intergranular lane. At short wavelengths the contribution functions are double peaked, at 0.4~mm the main peak is at the height of the temperature minimum with a secondary peak in the low chromosphere, whereas at 1~mm the main peak moves to chromospheric heights, but a small contribution from the temperature minimum region remains. At longer wavelengths the contribution heights move higher into the chromosphere, the contribution functions are sharply defined with a single peak close to the boundary between the chromosphere and the TR. The resulting mm brightnesses are significantly higher than the average values given in Table~\ref{table:1} and increase almost linearly with wavelength.
  \item Spatial location $(x,y)=(0.7,0.9)$~Mm (Fig.~\ref{fig7}, marked ''3'' in Fig.~\ref{fig5}): This location corresponds to the central part of the loop connecting the two main field concentrations and is characterized by  complicated temperature and density profiles showing a number of local enhancements along the line of sight. This leads to a compound contribution function at 3.6~mm. The brightness spectra achieve average values.
  \item Spatial location $(x,y)=(9.5,0.9)$~Mm (Fig.~\ref{fig8}, marked ''4'' in Fig.~\ref{fig5}): This is a peculiar location at the boundary between an intergranular lane and a granule. It is characterized by a bubble of very hot (basically coronal temperature), low-density gas at a height of about 2000~km, somewhat like a piece of corona, separated from the main corona. This feature, due to low electron density, produces low opacity and does not contribute to the emerging intensity. As a result, the contribution function at 3.6~mm has peaks on either side of this abrupt temperature enhancement. Radiation at 0.4~mm and 1~mm forms below these layers and is not affected by this inhomogeneity, while radiation at 10~mm is formed only above it. The brightness at 0.4~mm and 1~mm is typical for these wavelengths.  The $T_b$ at 3.6~mm and 10~mm exceeds the average values.

  \item Spatial location $(x,y)=(10.2,0.9)$~Mm (Fig.~\ref{fig9}, marked ''5'' in Fig.~\ref{fig5}): This is an example of the atmosphere above a granule. A steep transition region at 2000~km height results in a narrow contribution function for 10~mm and in an extremely low synthesized brightness. The radiation at shorter wavelengths receives its main contribution from the heights where the temperature is below 5000~K, and as expected, it also shows very low brightness. Both this location and the previous one display signs of a magnetic canopy with a base at around 500-700~km.
\end{itemize}

  \begin{figure}[!htb]
  \centering
           \includegraphics[trim=100 0 0 100,clip,width=0.65\textwidth]{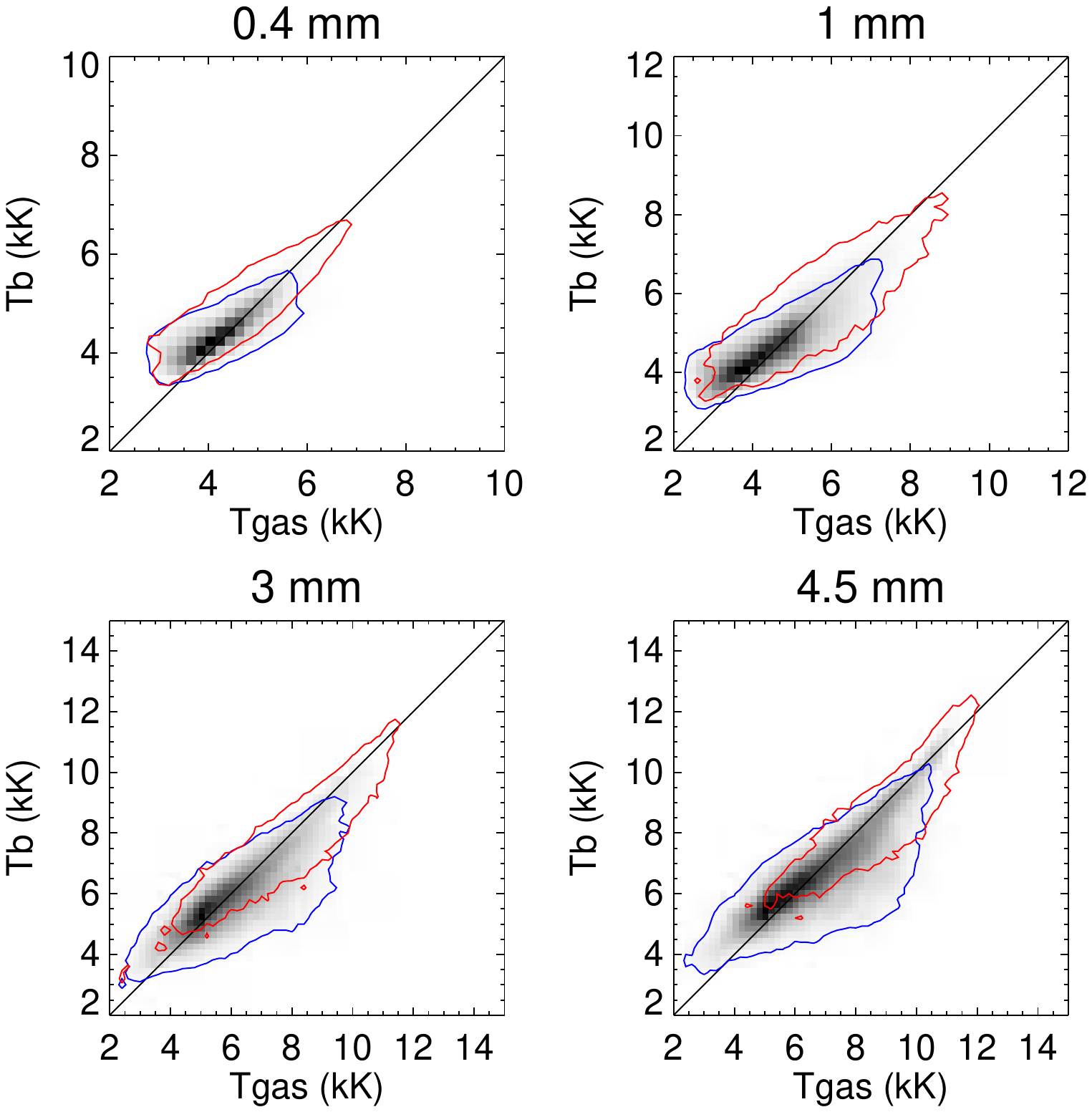}
      \caption{Density scatter plots (2D histograms) of the brightness temperature $T_b$ versus gas temperature $T_{gas}$ taken at the effective formation heights for 0.4~mm, 1~mm, 3~mm, and 4.5~mm. Darker color indicates more pixels in the bin. Solid lines denote $T_b=T_{gas}$. Blue contour (dark gray in the printed version) encloses 95\% of the spatial locations with weak magnetic field ($|B|<30$~G at the effective formation height), red contour (light gray in the printed version) includes 95\% of the locations of strong magnetic field ($|B|\geq30$~G at the effective formation height).
                    }
         \label{fig10}
   \end{figure}

\begin{figure}[!htb]
  \centering
            \includegraphics[width=0.21\textwidth,angle=90]{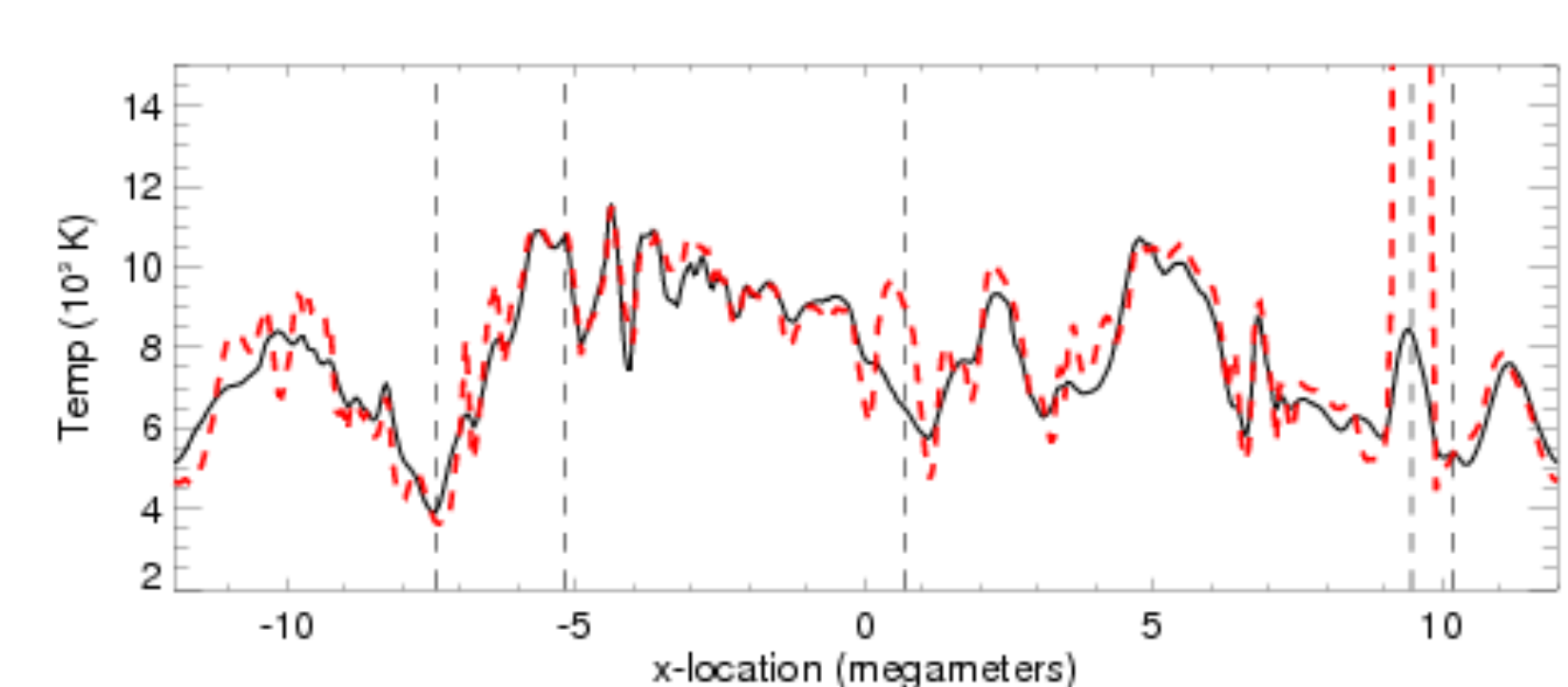}
      \caption{Brightness temperature at 3.6~mm (solid), electron temperature at the effective formation height (dashed) along the X-profile at Y=0.9~Mm. The vertical dashed lines are the same as in Figs.~\ref{fig3} and ~\ref{fig5}.
              }
         \label{fig4}
   \end{figure}

\subsection{Thermal diagnostics}\label{sectCor}

In this section we investigate the potential of submm/mm brightness spectra to diagnose chromospheric plasma. We present correlations between  brightness temperature and the properties of the atmosphere, in particular its thermal structure.

In Fig.~\ref{fig10} we correlate the synthetic mm brightness with the model temperature taken at the heights corresponding to the effective formation heights at wavelengths 0.4~mm, 1~mm, 3~mm, and 4.5~mm. Because of the many pixels in the snapshot, we show the correlations by binning the data into 2D histograms. In the density scatter plots we distinguish between a weak magnetic field with $|B|<30$~G at the effective formation height (darker gray contours in the printed version and blue contours in the online version), and an enhanced magnetic field with $|B|\geq30$~G (light gray contours in the printed version, red contours in the online version). The scatter seen in Fig.~\ref{fig10} illustrates the influence of the extended formation
height ranges, which instead of sampling the temperature at a fixed geometrical height, effectively mixes contributions from different layers.

For regions with large $B$, the largest differences between $T_{gas}$ and $T_b$ occur at the high and low end of the temperature range sampled at each $\lambda$. In the cool part of the range mm brightness tends to overestimate $T_{gas}$, while when $T_{gas}$ is high the synthetic brightness underestimates it. For weak magnetic features a wider ''scatter rectangle'' around the $T_{gas}=T_b$~curve in the scatter plots is seen. When confining the analysis to fields with $|B|\geq30$~G, the scatter becomes much smaller and mm brightness reproduces temperature at different chromospheric heights relatively well. The agreement for $|B|\geq30$~G is better for longer wavelengths, which has two reasons: applying a 30~G magnetic field strength threshold at the formation height means that we consider fewer regions at longer $\lambda$, since on average $B$ decreases with height. In the more strongly magnetized parts of the atmosphere contribution functions at longer wavelength show a simple form with one peak, see Fig.~\ref{fig6}, so that the effective formation height provides a good measure of the real formation height.

The relation between brightness temperature and atmospheric temperature can also be seen in Fig.~\ref{fig4}, where we
show brightness at 3.6~mm (solid), the longest of the currently available ALMA wavelengths, as a function of position along the analyzed XZ-cut, together with the model temperature (dashed) taken at the heights corresponding to the effective formation heights. In the $T_{gas}$ profile there are excursions to values lower and higher than the synthesized $T_b$. Furthermore, at $X=+9.5$~Mm,  there is a huge discrepancy between the two quantities. It corresponds to an intergranule-granule boundary. This location was analyzed in Fig.~\ref{fig8}. At this location the effective height evaluated from the centroid of the contribution function does not reflect the real formation heights because of the complex form of the CF that has two peaks of similar strength with negligible contribution from in between. For strong magnetic elements (e.g., spatial location $X=-5.2$~Mm) the agreement between synthetic brightness and atmospheric temperature is nearly perfect.

     \begin{figure*}[!htb]
  \centering
           \includegraphics[trim=100 0 100 200,clip,width=0.85\textwidth]{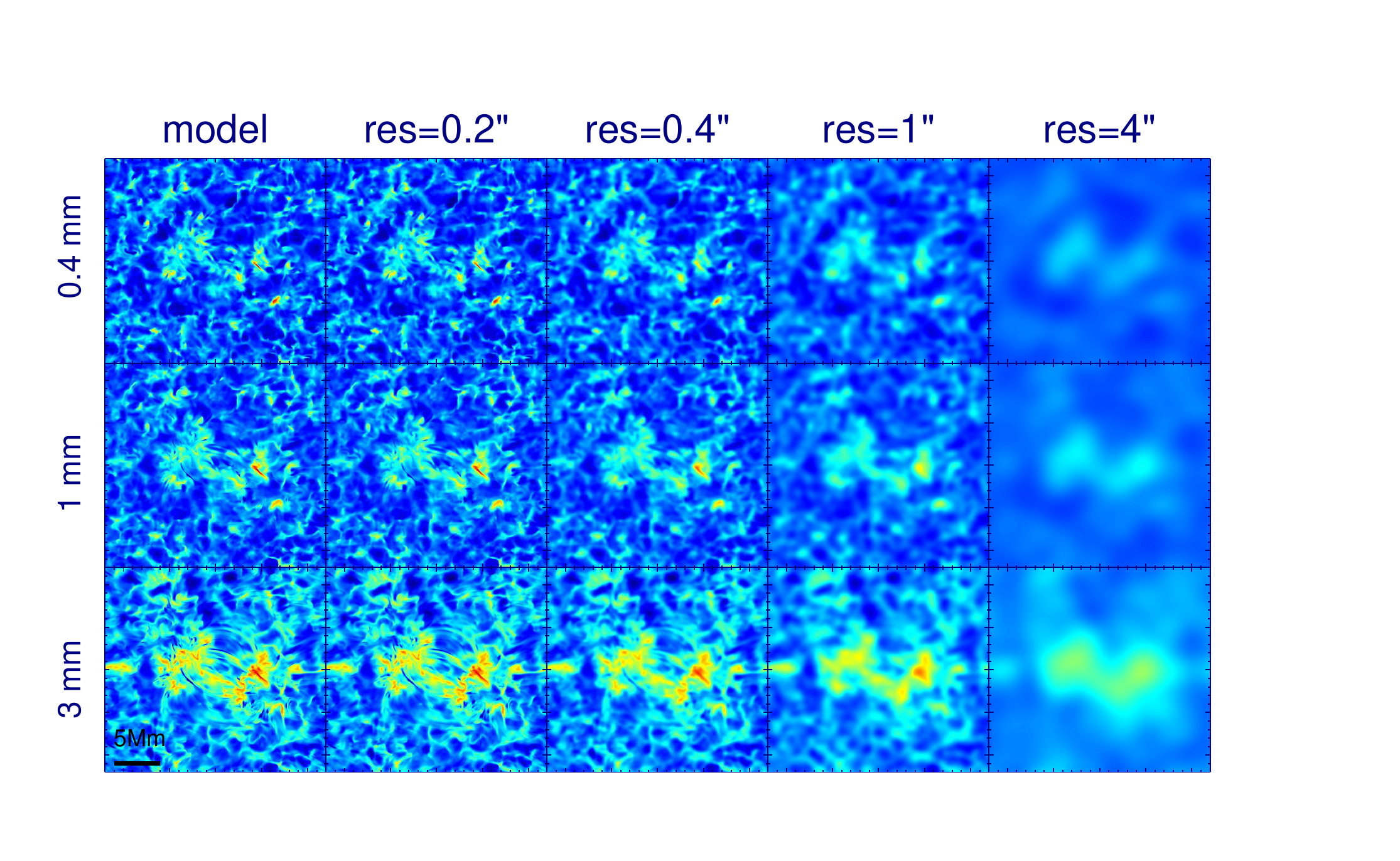}
      \caption{Spatially smeared synthetic brightness at 0.4~mm (top row), 1~mm (middle row), and 3~mm (bottom row) at a spatial resolution of 0.2\arcsec, 0.4\arcsec, 1\arcsec\ and 4\arcsec. The same color scale is used for all panels in a row. The field size is 24~Mm x 24~Mm.
              }
         \label{fig15}
   \end{figure*}

    \begin{figure}[!htb]
  \centering
           \includegraphics[trim=0 0 200 100,clip,width=0.35\textwidth]{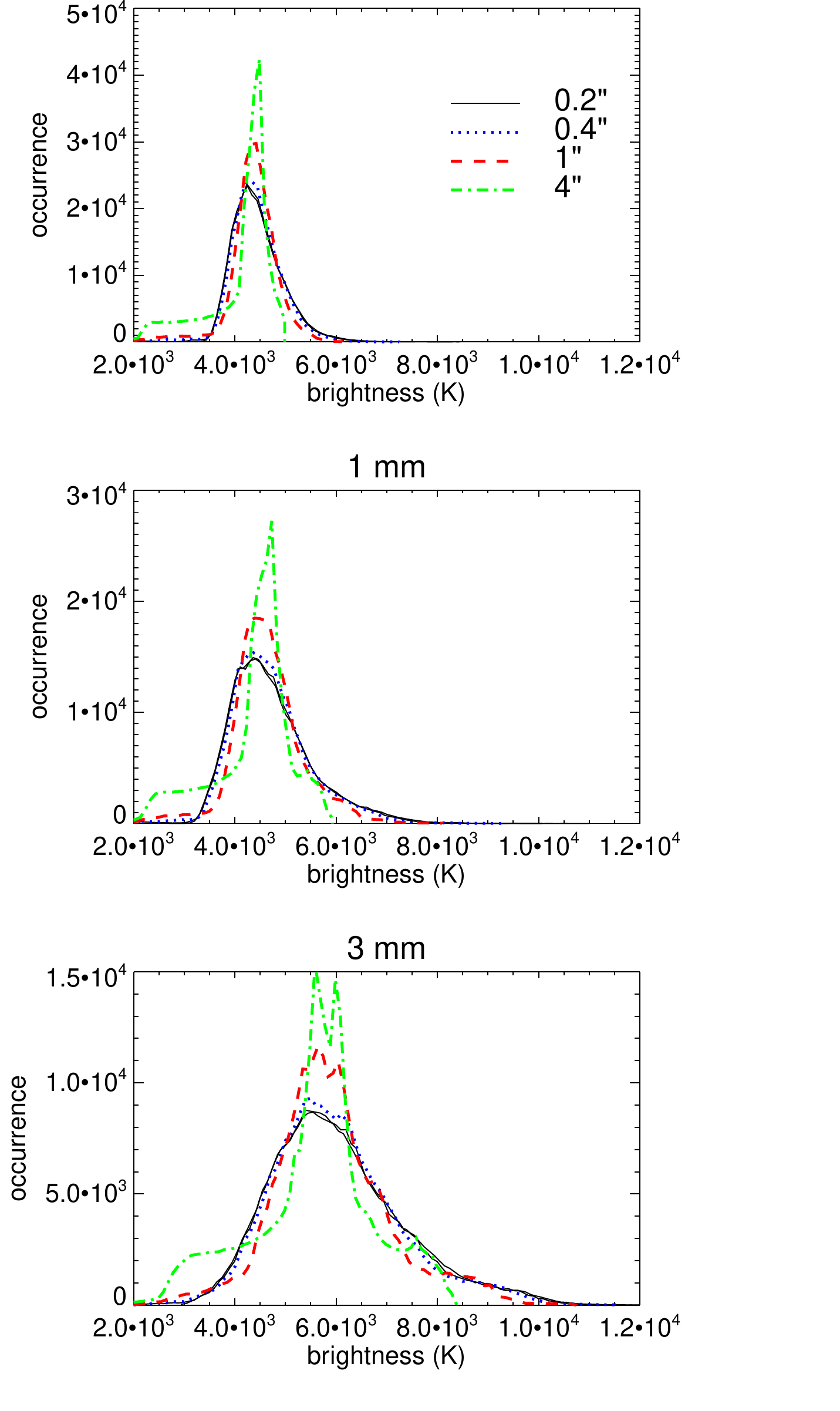}
      \caption{Spatial resolution effect on the brightness histogram. Top: brightness histograms at 0.4~mm for a spatial resolution of 0.2\arcsec\ (solid), 0.4\arcsec\ (dotted), 1\arcsec\ (dashed), and 4\arcsec\ (dot-dashed). Middle: the same for 1~mm. Bottom: the same for 3~mm.
              }
         \label{fig13}
   \end{figure}

    \begin{figure}[!htb]
  \centering
           \includegraphics[trim=100 0 0 100,clip,width=0.65\textwidth]{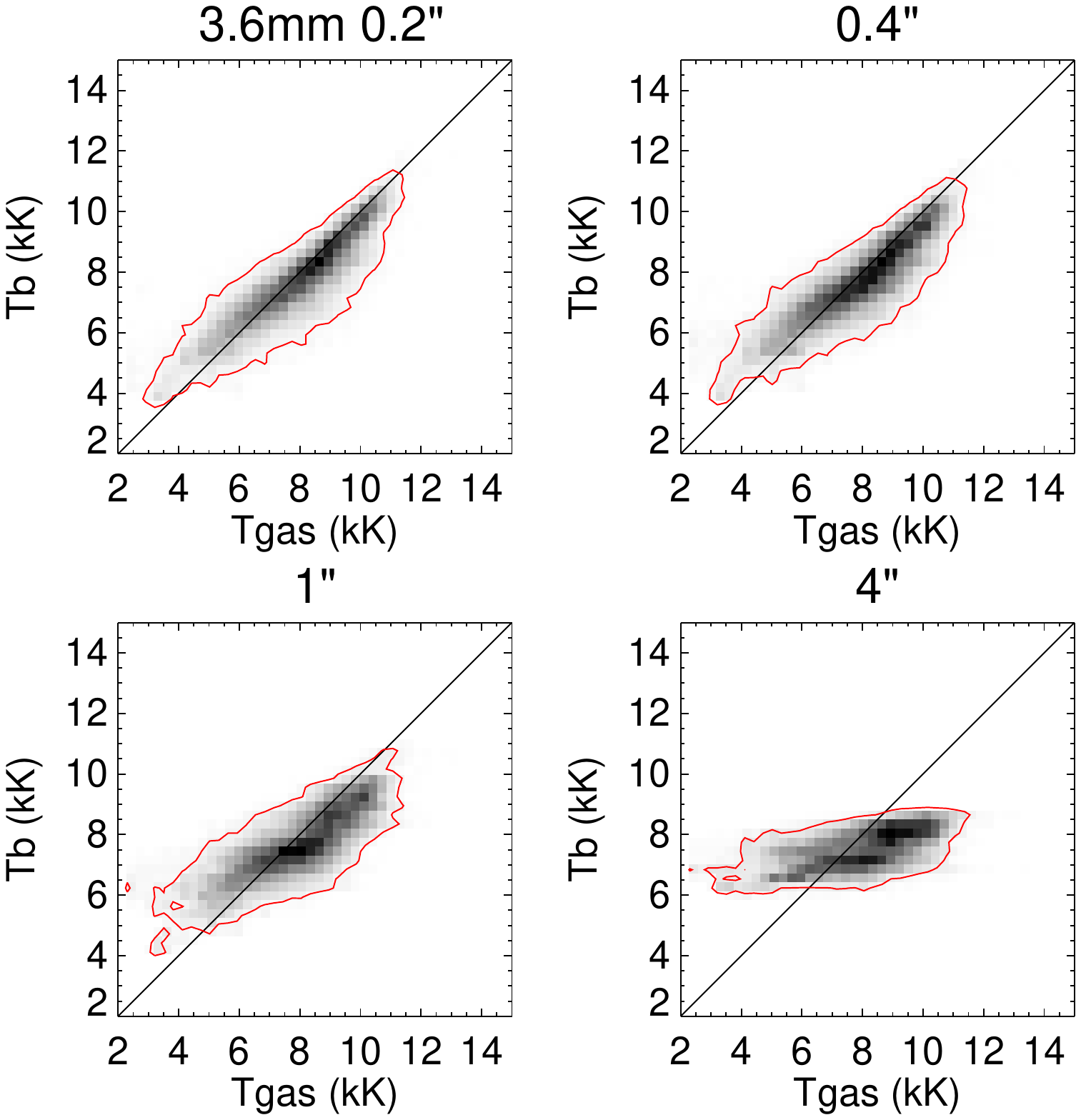}
      \caption{Density scatter plots of brightness temperature at 3.6~mm at resolutions of 0.2\arcsec, 0.4\arcsec, 1\arcsec\ , and 4\arcsec\ versus gas temperature $T_{gas}$ taken at the effective formation heights. Darker color indicates more pixels in the bin. The data are restricted to the locations of strong magnetic field ($|B|\geq30$~G) at the effective formation height. Red contours enclose 95\% of the selected pixels.
              }
         \label{fig14}
   \end{figure}

\subsection{Effect of spatial smearing}

Depending on array configuration (maximum baseline) and wavelength, ALMA can achieve a spatial resolution in the range 0.005\arcsec\ - 5\arcsec.
In Table~\ref{table:2} we list the estimates of the FWHM of the synthesized beam (point spread function), which is the inverse Fourier transform of a (weighted) $u-v$ sampling distribution. To account for this finite resolution, we spatially smeared the synthetic brightness maps by convolving with a Gaussian kernel of corresponding FWHM to mimic the instrumental profile. Note that we did not change the pixel size, so that in the following plots we oversample the spatially smeared data sets.

\begin{table}
\caption{Spatial resolution of the ALMA interferometer as a function of the longest array baseline for a number of wavelengths.}             
\label{table:2}      
\centering                          
\begin{tabular}{l c c }        
\hline\hline                 
\noalign{\smallskip}
$\lambda$, mm & FWHM(\arcsec) & FWHM(\arcsec) \\    
& maxbase$=150$~m & maxbase$=16$~km \\
\noalign{\smallskip}
\hline                        
\noalign{\smallskip}
0.4 & 0.5   &   0.005 \\
1.0 & 1.3   &   0.01 \\
3.0 & 4   &   0.04 \\
4.5 & 6    &  0.06\\
\noalign{\smallskip}
\hline                                   
\end{tabular}
\end{table}

The effect of spatial smearing on the mm brightness is illustrated in Fig.~\ref{fig15} for three wavelengths: 0.1~mm, 1~mm, and 3~mm, and for four values of FWHM: 0.2\arcsec, 0.4\arcsec, 1\arcsec\ , and 4\arcsec. While at 0.2\arcsec\ much of the fine structure of the original-resolution images is preserved, it is increasingly lost as the resolution is reduced further. At the same time, the intensity contrast is rapidly reduced, as can be deduced from the intensity histograms in Fig.~\ref{fig13} plotted for the same three wavelengths and from Table~\ref{table:3}. In Table~\ref{table:3} we list brightness variations $T_b^{rms}$ and relative brightness temperature contrast $\frac{T_b^{rms}}{<T_b>}$ as a function of ALMA resolution for comparison with the original values in Table~\ref{table:1}.

The difference between original brightness and brightness at a resolution of 0.2\arcsec\ is negligible. This is expected because
of the rather large grid spacing of 48~km of the original simulation, which makes the spatial resolution of the original simulation similar to the degraded one. At a resolution of 0.4\arcsec\ the finest structure starts to be washed out (but insignificantly), and the most extreme brightness values are no longer present in the histogram (see dotted curves in Fig.~\ref{fig13}). At all wavelengths the relative brightness contrast changes only
very little from its original values. However, there is a significant change at 1\arcsec\ resolution: the contrast begins to be reduced significantly (it decreases to 70\% of original contrast at the shortest wavelengths), which is visible as a narrowing of the intensity histogram (see dashed curves in Fig.~\ref{fig13}), the fine structure is already smeared out, while the larger scale pattern is well distinguishable and closely reflects that in the original image. Finally, at 4\arcsec\ resolution, which corresponds to the best resolution of the solar interferometric  observations with CARMA (Combined Array for Research in Millimeter-wave Astronomy) currently available, all the fine structure is lost and the contrast has dropped dramatically.

The good agreement between synthetic mm brightness and gas temperature evaluated at the effective formation height steadily degrades as the spatial resolution is decreased. In Fig~\ref{fig14} we show in the form of a 2D histogram how it develops for $\lambda$=3.6~mm from the original dependence to a spatial resolution of 4\arcsec\ for the computed $T_b$ maps. Note that the electron temperature at the effective formation height is not spatially degraded, since we wish to determine how much of the true local temperature can be deduced from degraded data.
 The range covered by $T_b$ decreases with decreasing resolution, the electron temperature does not, so that the relationship starts to deviate from the ideal line, mainly at low temperatures. Nonetheless, the diagnostics remains fairly robust up to 1\arcsec\ resolution, particularly for the somewhat higher temperatures. At 4\arcsec\ resolution, however, the observed $T_b$ carries almost no information on the original electron temperature in the fine-scale atmosphere.

\begin{table*}
\caption{RMS variation $T_b^{rms}$ and relative brightness temperature contrast $\frac{T_b^{rms}}{<T_b>}$ for a number of analyzed wavelengths as a function of the ALMA spatial resolution}             
\label{table:3}      
\centering                          
\begin{tabular}{l c c c c c c c c}        
\hline\hline                 
\\
Resolution  & \multicolumn{4}{c}{$T_b^{rms}$ (K)} & \multicolumn{4}{c}{$\frac{T_b^{rms}}{<T_b>}$} \\  
 (arc sec) & 0.4~mm & 1.0~mm & 3~mm & 4.5~mm & 0.4~mm & 1.0~mm & 3~mm & 4.5~mm\\
 \\
\hline                        
\\
\noalign{\smallskip}
0.2 & 496 & 811 & 1311 & 1465 & 0.11 & 0.17 & 0.22 & 0.22\\
0.4 & 442 & 738 & 1220 & 1369 & 0.10 & 0.15 & 0.20 & 0.20\\
1.0 & 339 & 591 & 1041 & 1192 & 0.08 & 0.12 & 0.17 & 0.18\\
4.0 & 153 & 310 & 667 & 819 & 0.03 & 0.06 & 0.11 & 0.12\\
\noalign{\smallskip}
\hline                                   
\end{tabular}
\end{table*}


\section{Discussion and conclusions}

Using realistic simulations of the quiet Sun, we studied simulated brightness at millimeter and submillimeter wavelengths and how it is affected
by the instrumental resolution of ALMA.

We conclude that millimeter brightness provides a reasonable measure of the chromospheric thermal structure at the height at which the radiation is formed. We found that the brightness temperature is a reasonable measure of the gas temperature at the effective formation height, defined as the height corresponding to the centroid of the contribution function, in spite of considerable scatter. The scatter is significantly reduced when very weak magnetic fields are avoided.
Our results indicate that although instrumental smearing reduces the correlation between brightness and temperature, $T_b$ can still be used to diagnose the electron temperature up to a resolution of 1\arcsec. Another effect of finite spatial resolution is that it smooths out the more extreme values.

When comparing images, spatial resolution emerges as a critical problem for chromospheric observations with ALMA. It will be important to achieve the highest possible resolution because the fine details are increasingly lost as the resolution decreases. We cannot conclude from the current simulation whether 0.1\arcsec\ is a sufficiently high resolution because of the rather large grid size of 48~km=0.064\arcsec\ of these simulations. Simulations with a finer grid may reveal more fine structure.

   \begin{figure*}[!htb]
  \centering
           \includegraphics[trim=0 0 0 300,clip,width=0.8\textwidth]{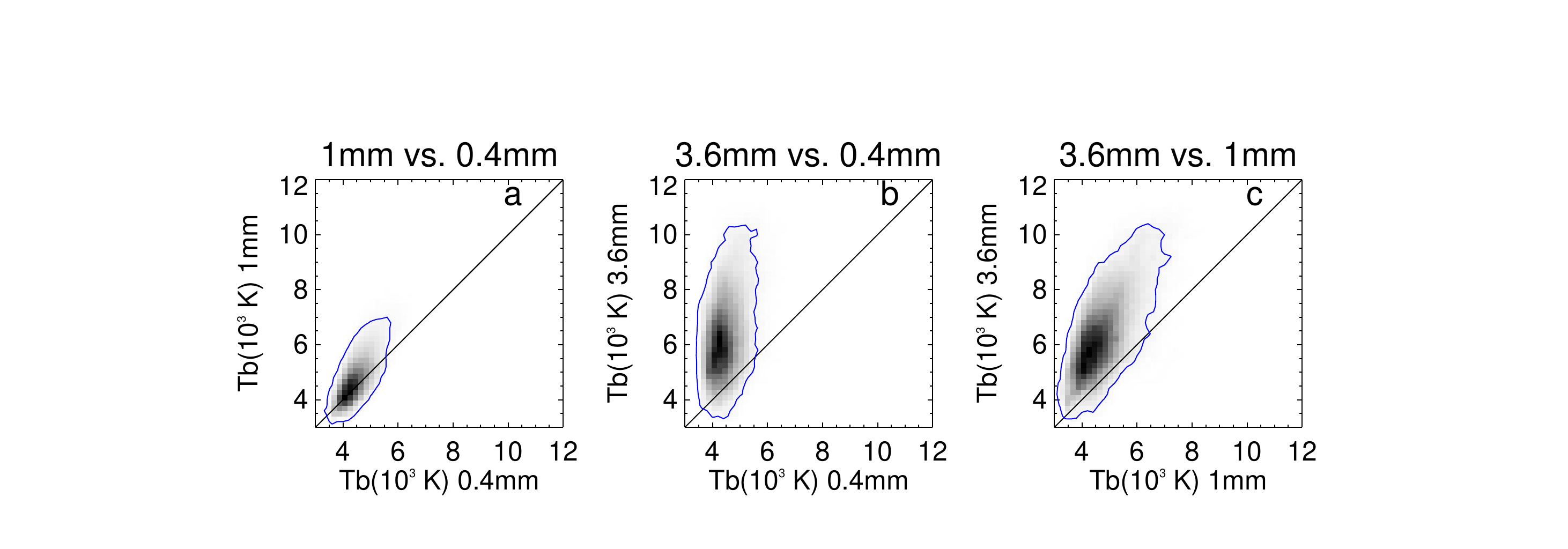}
      \caption{Density scatter plots of brightness temperatures $T_b$ for $\lambda$=0.4~mm, 1~mm, and 3.6~mm. Darker color indicates more pixels in the bin. Blue contours enclose 95\% of the pixels. Solid lines denote $T_b^{\lambda1}=T_b^{\lambda2}$.
              }
         \label{fig12}
   \end{figure*}

The often complex contribution functions of the mm radiation imply that the brightness temperatures represent the integrated physical state of an extended height range, sometimes also over two distinct, well-separated height ranges. However, multiwavelength observations with ALMA with a narrow step in wavelength can provide chromospheric images formed over subtly different height ranges. This will make tomography of the chromospheric thermal structure feasible (called ''volume imaging'' by \citet{wed07}). This conclusion is confirmed by Fig.~\ref{fig12}, which displays density scatter plots of brightness temperatures at three wavelengths corresponding to different ALMA frequency bands.  Neighboring bands (Figs.~\ref{fig12}a and~\ref{fig12}c) show less deviation from diagonal lines, depicting expected identical brightness for the two wavelengths and thus carry some overlapping information, in contrast to the bands farther away (Fig.~\ref{fig12}b), which show larger scatter.

ALMA will provide a unique opportunity to probe the solar atmosphere from the height of the classical temperature minimum through the chromosphere to just below the transition region (see Fig.~\ref{fig3}). The shortest currently available wavelengths at ALMA will examine the temperature minimum region in the solar atmosphere, while in the range from 1~mm to 3~mm we can gain access to the middle chromospheric heights (at 1000~km-2000~km). At the longest wavelengths, planned for the future development of ALMA, the transition region becomes accessible. Moreover, solar chromospheric science can profit from the ALMA observations at submillimeter and millimeter wavelengths used in combination with other chromospheric diagnostic such as Ca~{\sc ii}~K, Mg~{\sc ii}~k \citep{leenaarts13a,leenaarts13b,Danilovic,Riethmuller}, Ca~{\sc ii}~854.2 nm \citep{rodriguez}, or H$\alpha$ \citep{leenaarts12}.

As the next steps of this study, we are planning to study diagnostics of magnetic fields at the heights where submm-mm radiation is formed. We will discuss simulated circular polarization in the context of successful solar ALMA polarization measurements. We will reconsider the assumptions made in this work for computing the submm-mm radiation, in particular the use of quasilongitudinal approximation. Next we plan to go beyond a single snapshot to investigate the time dependence and the necessary time resolution of ALMA observations.

In this paper we used one of the most realistic simulations of the solar chromosphere currently available. However, as reported by \citet{leenaarts13b}, there are two processes of particular importance for the chromospheric thermal structure
that are not included in the numerical simulations analyzed here. These refer to the non-equilibrium ionization of
helium and the effect of partial ionization on chromospheric heating by magnetic fields \citep[see][and references therein]{leenaarts13b}. Studies based on simulations of regions with different amounts of magnetic flux would also be useful, as well as simulations with a smaller grid size.

\begin{acknowledgements}
We thank Andreas Lagg for computational support. This work was partly supported by the BK21 plus program through the National Research Foundation (NRF) funded by the Ministry of Education of Korea. The research leading to these results has received funding from the European Research Council under the European Union's Seventh Framework Programme (FP7/2007-2013) / ERC grant agreement nr. 291058 and from the Research Council of Norway through the project "Solar Atmospheric Modelling" and a grant of computing time from the Program for Supercomputing. M.Loukitcheva acknowledges Saint-Petersburg State University for a research grant 6.0.26.2010.

\end{acknowledgements}

\end{document}